\documentclass[12pt]{iopart}

\usepackage{graphicx}
\usepackage{subfigure,cite}
\usepackage{iopams}

\begin{document}

\title{Fidelity of the near resonant quantum kicked rotor}
\author{Probst B, Dubertrand R and Wimberger S}
\address{Institute for Theoretical Physics and Center for Quantum
  Dynamics, University of Heidelberg, Philosophenweg 19, D-69120
  Heidelberg} 
\begin{abstract}
We present a perturbative result for the temporal evolution of
the fidelity of the quantum kicked rotor, i.e. the overlap of the same
initial state evolved with two 
slightly different kicking strengths, for kicking periods close to a
principal quantum resonance. Based on a
pendulum approximation we describe the fidelity for rotational orbits
in the pseudo-classical phase space of a corresponding classical
map. Our results are compared to numerical simulations indicating the
range of applicability of our analytical approximation. 
\end{abstract}

\pacs{03.65.Sq, 05.60.Gg, 03.75.Dg, 37.10.Vz} 

\section{Introduction}

In classical mechanics chaos can be defined using the stability of a
trajectory. Consider two neighbouring phase
space points, the distance of their orbits in the phase space will
grow exponentially in time for a generic chaotic system.
Such an approach is bound to fail in a quantum mechanical
treatment as the time evolution is unitary and the overlap of two wave
packets is constant in time. However changing a parameter of the Hamiltonian
instead of the initial state will lead to an overlap varying in
time. This idea was formulated by Peres \cite{peres} when he 
introduced the fidelity, also known as Loschmidt echo
\cite{JP2001}. The fidelity is defined 
as the overlap of an 
initial state evolved with slightly different Hamiltonians. This
quantity has been used to characterise the stability of quantum
states \cite{Gorin2006}. For classically chaotic systems, the fidelity shows
some generic behaviour \cite{Gorin2006}. 

Here we study the fidelity in the quasi-integrable regime for which not
so many results are established
\cite{Casati2003,Sankaranarayanan2003,Abb2009,Fishman2011,Weinstein2005}. We
will do this for one of the best known examples of classically chaotic system:
the kicked rotor (KR) \cite{Chirikov1979,Lichtenberg1992}. In order to
understand the fingerprints of classical chaos in quantum mechanics the quantum
kicked rotor (QKR) has been and still is a fruitful field of study
\cite{Izrailev1990}. It shows several interesting phenomena like
quantum resonance \cite{Izrailev1980} and dynamical localisation
\cite{Fishman1982} both in direct contradiction to the behaviour
of the classical system.  
In the implementation of the QKR near a quantum resonance in the
gravity field Oberthaler {\em et al.} observed quantum
accelerator modes \cite{Oberthaler1999}. Fishman {\em et al.} were able to describe these
theoretically using a pseudo-classical limit. There  the detuning of
the kicking period to its resonant value 
plays the role of the Planck constant \cite{Fishman2003}. Using this treatment the QKR
can be mapped onto a kicked rotor with a renormalized kicking strength. This
leads to regular structures and allows the application of
semi-classical methods in the pseudo-classical limit although the
system might be chaotic in the true semi-classical limit \cite{WGF2003}. 

The KR shows two types of motion in the quasi-integrable regime,
oscillations about the stable fixed points and rotating motion. The motion on the classical resonance island surrounding the 
fixed point leads to revivals of the quantum fidelity. This was shown for
the regular QKR by Sankaranarayanan  {\em et al.} \cite{Sankaranarayanan2003}
and for the near resonant QKR by Abb  {\em et al.} \cite{Abb2009}. Similar
results were obtained for another kicked system by Krivolapov {\em et al.}
\cite{Fishman2011}. Rotational and oscillating orbits were also
numerically studied in \cite{Weinstein2005}. Our focus in this paper is to treat the rotating modes
of the QKR near a quantum resonance. 
Therefore we will use the pseudo-classical method and apply the pendulum approximation. Section \ref{sec:FidInAOKR} sets the 
stage by reviewing the pseudo-classical approximation and defining the fidelity which is studied here.
In Section \ref{sec:PertTreatment} we will give a perturbative treatment of the
pendulum and discuss its validity based on a comparison with numerical simulations in Section \ref{sec:numerics}. 
An additional numerical check is shown in Sect.~5.

\section{Fidelity for the atom optical kicked rotor}
\label{sec:FidInAOKR}

The story of the experimental investigation of the quantum kicked
rotor is quite long \cite{Raizen1995} and it has been continuing until
today (see, e.g., the refs. \cite{theory_many_part,expBEC,exp_now} and, for fidelity
measurements specifically, the refs. \cite{exp_fid}). In the
experimental realisations, atoms are kicked by a periodic potential
formed by a standing wave of laser light, i.e. an optical lattice
which is flashed on and off periodically in time
\cite{GSZ1992,Raizen1995}. Using rescaled dimensionless momentum $p$,
position $x$, kicking period $\tau$ and kicking strength 
$k$, see e.g. \cite{Fishman2003}, the Hamiltonian for one kicked atom is:
\begin{equation}
 H(p,x,t)=\frac{p^2}{2}+k\cos(x)\sum_{n=-\infty}^\infty\delta(t-n\tau).
\end{equation}
The Floquet operator mapping the state right after one 
kick to the state right after the next kick is, see e.g. \cite{Izrailev1990},
\begin{equation}
\hat \mathcal{U}=\e^{-\rmi k\cos(\hat X)}\e^{-\rmi\frac{\tau}{2}\hat P^2}.
\end{equation}
The dynamics is obtained by repeated application of
this operator. In contrast to the usual kicked rotor, a kicked atom
lives along a line. Doing a gauge transformation
one can still recover a problem with conserved quasi-momentum, see
e.g. \cite{Fishman2003,WGF2003}. Here the quasi-momentum corresponds to the 
fractional part of the momentum $p=n+\beta$, where $n$ is an integer
and $\beta$ is a real variable between $0$ and $1$. The problem of one
atom along a line is mapped onto a problem of a continuous family of
rotors. Each of them corresponds to one value of $\beta$. That is the
reason why we will speak from now on of a $\beta-$rotor. 
In realisations of the kicked rotor with Bose-Einstein condensates it has been checked that
the interactions between atoms in the cloud can be neglected
\cite{theory_many_part,expBEC}. This brings another justification of our one particle approach.
The wave function
of a single $\beta-$rotor $| \Psi_\beta\rangle$ is obtained from the wave function of the
kicked atom $| \psi\rangle$ by \cite{Fishman2003,WGF2003}:
\begin{equation}
 \langle \theta|\Psi_\beta\rangle=\frac{1}{\sqrt{2\pi}}\sum_{n\in\mathbb{Z}}\langle
 n+\beta|\psi\rangle\e^{\rmi n\theta}, 
\end{equation}
while the Floquet operator for one $\beta-$rotor is 
\begin{equation}
 \hat \mathcal{U}_\beta=\e^{-\rmi k\cos\hat\theta}\e^{-\rmi\frac{\tau}{2}(\hat\mathcal{N}+\beta)^2},\label{eq:FloquetOperatorBloch}
\end{equation}
where $\hat\mathcal{N}$ is the
angular momentum operator. The operator
(\ref{eq:FloquetOperatorBloch}) formally differs from the usual Floquet
operator \cite{Izrailev1990} only by the quasi-momentum $\beta$. 

A principal quantum resonance occurs in the kicked particle when the phase due to the free evolution in
(\ref{eq:FloquetOperatorBloch}) vanishes \cite{Izrailev1986}. Each
resonance leads to a ballistic motion and a quadratic growth of the
energy. This happens at $\tau=2\pi l$ and for resonant quasi-momenta $\beta_{\rm res}=1/2+q/l$ for integers $l$
and $q$ such that $l\ge 1$ and  $0\le q \le l-1$. To simplify the
notation, in this paper we consider the specific resonance 
with $\tau=2\pi$ and $\beta_{\rm res}=1/2$.  

\begin{figure}
 \centering
 \includegraphics[clip]{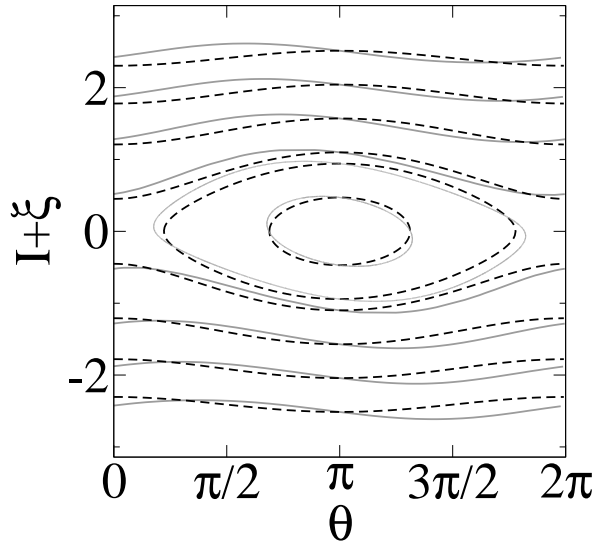}
  \caption{(Colour online) Phasespace of the pendulum (black \dashed) and
    the kicked rotor (grey \full) for $\tilde{k}=|\epsilon| k=0.08\pi$.}
  \label{fig:phasespacePenKR}
\end{figure}

Take now a kicking period slightly detuned from its resonant value
$\tau=2\pi +\epsilon$. 
Introducing the rescaled momentum $\hat I=|\epsilon|\hat{\mathcal{N}}$
we can rewrite the Floquet operator \eref{eq:FloquetOperatorBloch} as 
\begin{equation}
 \hat {\cal U}_{k,\beta}=\e^{-\frac{\rmi}{|\epsilon|}\tilde
   k\cos(\hat\theta)}\e^{-\frac{\rmi}{|\epsilon|}\left[\mathrm{sgn}(\epsilon)
   \frac{\hat{I}^2}{2}+\hat I(-\pi +\tau\beta)\right]} \label{U_beta_r}\ .
\end{equation}
This operator can be identified as the formal quantisation of another
\emph{fictitious} kicked rotor. The main benefit of this mapping is that 
the kicking strength of the new problem  is $\tilde k=|\epsilon| k$
and $|\epsilon|$ plays the role of the Planck constant. We will
be interested from now on in the regime $|\epsilon|\to 0$, called
$\epsilon-$semiclassical limit \cite{Fishman2003}, which must not be confused with the
\emph{true} semiclassical one of the initial problem. In the
$\epsilon-$semiclassical limit one can derive easily the $\epsilon-$classical map,
which is very similar to the celebrated standard map \cite{Chirikov1979}:
\begin{eqnarray}
 I_{t+1}=I_t+\tilde{k}\sin(\theta_{t+1})\nonumber\\
 \theta_{t+1}=\theta_t+\mathrm{sgn}(\epsilon) I_t -\pi +\tau \beta \
 \ \textrm{mod}\ 2\pi \ \label{EpsClassStandardMap}.
\end{eqnarray} 
We are interested in the dynamics around a stable fixed point of
(\ref{EpsClassStandardMap}). One common approximation is the pendulum
approximation \cite{Chirikov1979,Lichtenberg1992}: 
\begin{equation}
 H_\mathrm{Pen}(I,\theta)=\frac{(I+\xi)^2}{2}+\tilde k\cos\theta
\label{Hpendulum} \ ,
\end{equation}
where we have defined $\xi=\textrm{sgn}(\epsilon)(-\pi +\tau\beta)$.
For a given (even large) kicking strength $k$, one can always choose a
small enough $\epsilon$ so that we are in the quasi-integrable regime.
This makes the phase
spaces corresponding to (\ref{EpsClassStandardMap}) and
(\ref{Hpendulum}), respectively, very similar, see Fig.~\ref{fig:phasespacePenKR}.

The main goal of this paper is to check the stability of the quantum
dynamics under slight variation of the kicking strength. The fidelity \cite{peres,Gorin2006} 
appears as the natural quantity to look at. 
For one single $\beta-$rotor it is
defined as:
\begin{equation}
 F_\beta(k_1,k_2,t)=\Big|\langle \psi_0|{\mathcal{U}^{t\
     \dagger}_{k_1,\beta}}\ \mathcal{U}_{k_2,\beta}^t|\psi_0\rangle\Big|^2.\label{Fid1rotor}
\end{equation}
For the initial problem of a kicked atom one needs to consider the
fidelity for a sub-ensemble of rotors, which is defined then by \cite{Wimberger2006,Wimberger2004}:
\begin{equation}
 F(k_1,k_2,\beta_1,\Delta \beta,t)=\bigg|\int_{\beta_1}^{\beta_1+\Delta\beta}\rmd\beta \ \langle \psi_0|{\mathcal{U}^{t\
     \dagger}_{k_1,\beta}}\ \mathcal{U}_{k_2,\beta}^t|\psi_0\rangle\bigg|^2 \,. \label{eq:DefEnsemble}
\end{equation}
If the initial state mainly lives on the stable island the fidelity
shows revivals as explained in \cite{Abb2009}, see also a similar
context in \cite{Fishman2011}. Here the
study will be devoted to the fidelity (\ref{Fid1rotor}) in the neighbourhood of
such an island. 
We will approximate (\ref{Fid1rotor}) by the fidelity of the pendulum
(\ref{Hpendulum}). The fidelity of the pendulum is obtained by
expanding the initial state in the eigenbasis $|\phi_{n}(k)\rangle$
of the Hamiltonian\footnote{We emphasise the dependence of the
  eigenstates on $k$ as it is the perturbation parameter in the
  fidelity.} of Eq.~(\ref{Hpendulum}), which depends on $\beta$ 
via $\xi$. 
\begin{eqnarray}
F_\beta(k_1,k_2,t)=&\nonumber\\
\fl \bigg|\sum_{n,m}\langle\Psi_\beta(t=0)|\phi_{n}(k_2)\rangle \langle
 \phi_{n}(k_2)|\phi_{m}(k_1)\rangle \langle\phi_{m}(k_1)
 |\Psi_\beta(t=0)\rangle\e^{\rmi\frac{t}{|\epsilon|}(E^{k_2}_n-E^{k_1}_m)}\bigg|^2&\
 . \label{DefFidpendulum}
\end{eqnarray}
Throughout the paper we focus on the quantum problem associated to
\eref{U_beta_r} so that $|\epsilon|$ is always our Planck
constant. This is the reason why we will call the
$\epsilon-$semiclassical regime simply the semiclassical regime. 

\section{Perturbative treatment of the pendulum}\label{sec:PertTreatment}

We are interested in the quantum pendulum following the approximation
(\ref{Hpendulum}). It is well known that this system is quantum
mechanically integrable \cite{condon}. Following
(\ref{DefFidpendulum}) we want to obtain simple explicit formul\ae{}
for the eigenenergies $E_n^k$ and the eigenfunctions
$|\phi_n(k)\rangle$ of the Hamiltonian when $|\epsilon|$ is going to
$0$. We will follow standard perturbation theory. The only unusual
thing is that the potential is proportional to the effective Planck
constant so that it vanishes at the classical limit $|\epsilon|=0$. 

\begin{figure}[!ht]
 \centering
 \includegraphics{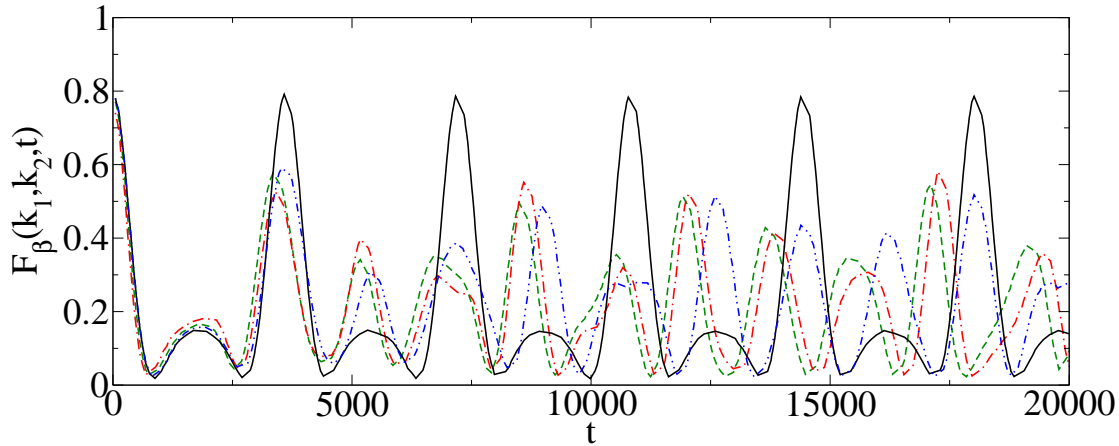}
 \caption{(Colour online) Fidelity using the pendulum (green \dashed),
   the perturbative result
   with the third (black \full) and fourth (blue \dashddot) order in $|\epsilon|$ in
   the energy, and the original QKR (red \chain). $\beta=0.3$, $\epsilon=0.05$,
   $k_1=0.6\pi$ and $k_2=0.8\pi$. The data are averaged over $100$
   kicks in order to cancel fast oscillations.} 
 \label{fig:QualPert}
\end{figure}

Our unperturbed system is a free particle along a ring with
eigenenergy and eigenfunction ($m\in\mathbb{Z}$):
\begin{eqnarray}
E_m&=&\frac{(m|\epsilon|+\xi)^2}{2}=\frac{\xi^2}{2}+\xi
m|\epsilon|+{\cal O}(|\epsilon|^2)\ ,\label{E_unpert}\\
\langle\theta| \phi_m\rangle &=&\frac{e^{\rmi m
    \theta}}{\sqrt{2\pi}}\ . \label{phi_unpert}
\end{eqnarray}
For $\tilde{k}>0$ the Schr\"odinger equation for the stationary states
becomes: 
\begin{equation}
\frac{1}{2}\left(-\rmi|\epsilon|\frac{\partial}{\partial \theta}
+\xi\right)^2\Psi + \tilde{k}\cos\theta\Psi=E\Psi \ .
\label{schro}
\end{equation}
Doing the gauge transformation
$\Psi=\exp(-\rmi\xi\theta/|\epsilon|) \psi$ and setting
$\psi(\theta)=f(z=\theta/2)$, Eq.~(\ref{schro}) becomes:
\begin{equation}
\frac{{\rm d}^2 f}{{\rm d} z^2} +\left(\frac{8E}{|\epsilon|^2} -
2\frac{4\tilde{k}}{|\epsilon|^2}\cos(2z)\right)f(z)=0 \ ,
\label{mathieu}
\end{equation}
which is the standard form of Mathieu equation, see e.g. 16.2.1 p.97
in \cite{bateman3}. For our purpose it is easier to look for a solution of
(\ref{mathieu}) as the following series:
\begin{equation}
f(z)=\sum_{n\in\mathbb{Z}} c_n e^{(\mu+2\rmi n)z} \ .
\label{hill}
\end{equation}
As we require the ``true'' wave function $\Psi(\theta)$ to be
univalued one can easily see that we need
\begin{equation}
\mu=\frac{2\rmi\xi}{|\epsilon|}\ .
\end{equation}
The eigenenergies of (\ref{schro}) are given by characteristic values of
Mathieu functions, which do not lead to simple explicit
formul\ae{}. A 
semiclassical approach is rather used to write an
expansion of the eigenenergies. The details are found in the
\ref{app_WKB}. The results are, assuming $\tilde{k}=k|\epsilon|$ and noting
$\xi_0=\textrm{sgn}(\epsilon)\pi(2\beta-1)$: 
\begin{eqnarray}
  E_m^k&\simeq&\frac{\xi_0^2}{2}+\xi_0 (m+\beta)|\epsilon|+
  \left(\frac{(m+\beta)^2}{2}+ \frac{k^2}{4\xi_0^2}\right)
  |\epsilon|^2-\frac{(m+\beta) k^2}{2\xi_0^{3}}|\epsilon|^3
  \nonumber\\ &&
  +\left(\frac{3}{4}\frac{(m+\beta)^2 k^2}{\xi_0^4}
    +\frac{5}{64}\frac{k^4}{\xi_0^6}\right)|\epsilon|^4\ .
  \label{E_pert}
\end{eqnarray}
In (\ref{hill}) the coefficients $c_n$ are the solutions of the following
recurrence relation:
\begin{equation}
\left[ 2E-(n|\epsilon|+\xi)^2\right] c_n=\tilde{k}(c_{n-1}+c_{n+1})\ .
\label{relcn}
\end{equation}
If we assume now that we start from an unperturbed state
(\ref{phi_unpert}) with the energy (\ref{E_unpert}), (\ref{relcn}) can be rewritten as: 
\begin{equation}
  \label{newrelcn}
  2\frac{(n-m)\xi_0}{-k}c_n^{(m)}=c_{n-1}^{(m)}+c_{n+1}^{(m)}\ ,
\end{equation}
which gives the solution: $c_n^{(m)}=J_{n-m}(-k/\xi_0)$ where $J_n(x)$
stands for the Bessel function of integer order $n$. The perturbed
eigenfunctions are then:
\begin{equation}
  \label{phi_pert}
  \langle \theta|\phi_m(k)\rangle =\sum_{n\in\mathbb{Z}}
  J_{n-m}\left(\frac{-k}{\xi_0}\right) e^{\rmi n\theta}=e^{\rmi
    m\theta-\rmi k\sin(\theta)/\xi_0}
\end{equation}
where we have used in the second equality the following identity for the Bessel
functions, see e.g. 7.2.4(26) p.7 in \cite{bateman2}:
\begin{equation}
  \label{FgenJ}
  \sum_{n\in\mathbb{Z}} J_n(x)e^{\rmi n\theta}=e^{\rmi
    x\sin\theta}\ .
\end{equation}
The great benefit from (\ref{phi_pert}) is that we can directly
compute the overlap coefficient for the fidelity
(\ref{DefFidpendulum}). We assume that the initial state is a plane wave
with momentum $n_0$: $\langle n|\Psi_\beta(t=0)\rangle=\delta_{n,n_0}$ where
$\delta_{n,k}$ is the Kronecker symbol. Then:
\begin{eqnarray}
  \label{overlap1}
  \langle \phi_m(k)|\Psi_\beta(t=0)\rangle&=\sum_{n\in\mathbb{Z}} \langle
  \phi_m(k)|n\rangle \langle n
  |\psi_\beta(t=0)\rangle=&
  J_{n_0-m}\left(\frac{-k}{\xi_0}\right)\ ,\\
\langle \phi_m(k_1)|\phi_n(k_2)\rangle&= \sum_{p\in\mathbb{Z}} \langle
  \phi_m(k_1)|p\rangle \langle p|\phi_n(k_2)\rangle
  \nonumber\\&=
  J_{m-n}\left(-\frac{k_2-k_1}{\xi_0}\right) \ .&
\label{overlap2}
\end{eqnarray}
Finally our simple perturbative approach lets us write an explicit
formula for the fidelity, reminding $\xi_0=\textrm{sgn}(\epsilon)\pi(2\beta-1)$:
\begin{eqnarray}
   F_\beta(k_1,k_2,t)=&\nonumber\\ \bigg|\sum_{n,m\in\mathbb{Z}}
   J_{m-n}\left(\frac{k_1-k_2}{\xi_0}\right)&
   J_{n_0-m}\left(\frac{-k_1}{\xi_0}\right)
   J_{n_0-n}\left(\frac{-k_2}{\xi_0}\right) e^{\rmi\frac{t}{|\epsilon|}(E^{k_2}_n-E^{k_1}_m)}\bigg|^2 \, , 
\label{Fidpend}
\end{eqnarray}
where $E_m^{k_1}$ and $E_n^{k_2}$ are given by
(\ref{E_pert}). {The formula (\ref{Fidpend}) is the main result
  of this paper. }
In the next section we will estimate numerically the range of validity and the
accuracy of (\ref{Fidpend}) to describe the quantum kicked rotor.
In Fig.~\ref{fig:QualPert} we can already see that the more orders we
take for the energy, the more accurate we get.

\section{Numerical comparison of the approaches}\label{sec:numerics}

The perturbative approach, cf. Eq.~\eref{Fidpend}, will now be checked numerically in this
section. As single rotors and ensembles show qualitatively different
behaviour we treat these
cases separately. 

Before showing the main results, we discuss the intrinsic limitations
of our approach. The phase space of the pendulum is the cylinder
whereas the phase space of the KR can be mapped onto a torus. The
pendulum approximation is therefore only valid in one phase space cell of
the KR. Due to this mismatch we expect the approximation to fail at
the border of a cell. 

In the derivation of the perturbative result we explicitly focused on
the rotating regime, which means that our results have to fail in the
description of states on the island. The half width of the island in the
pendulum approximation is given by $\Delta I=2\sqrt{k|\epsilon|}$
\cite{Lichtenberg1992}, which becomes in
units of the quasi-momentum  
\begin{equation}
 \Delta \beta_c^\mathrm{th}=\frac{\Delta
   I}{\tau}\approx\sqrt{\frac{|\epsilon|
     k}{\pi^2}}  \ \label{eq:Estimation_Island}. 
\end{equation}
When the distance from $\beta$ to its resonant value (centred at the
island) is less than
$\Delta\beta_c$ we expect to observe the revivals in the fidelity as
described in \cite{Abb2009}. 

\subsection{Single rotors}
\label{singlerotor}
\begin{figure}
 \centering
 \includegraphics{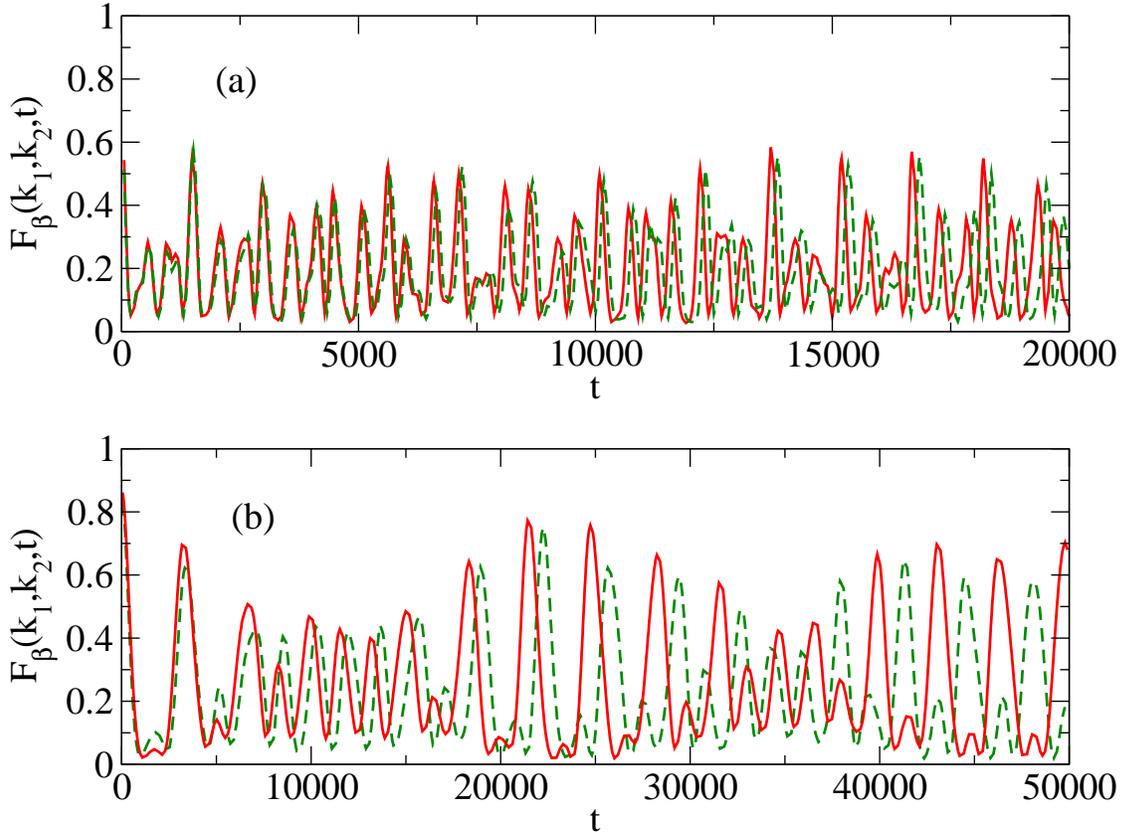}
 \caption{(Colour online) Fidelity for the QKR (red~\full) and the
   pendulum (green~\dashed) for  $\epsilon=0.075$, $k_1=0.6\pi$ and
   $k_2=0.8\pi$. (a) $\beta=0.3216$, (b) $\beta=0.2412$. The data are
   averaged over $100$ kicks in order to get rid of the fast oscillations.} 
 \label{fig:definition_of_criteria_singlerot}
\end{figure}

The most important goal of the discussion of single rotors is to get an
intuition for the quality of the pendulum approximation. We can
discuss this step only for single rotors as the calculation of the
pendulum fidelity is numerically very challenging. 
Averaging the fidelity over $100$ kicks allows to identify maxima in
the two cases and to read of the amplitude and the period 
(c.f., for instance, Fig.~\ref{fig:definition_of_criteria_singlerot}). Plotting the relative
deviation of the period of the maxima one observes that this relative
error is nearly independent of the choice of the maximum. As a measure
of deviation of the amplitude we compared a limited number of
maxima, whilst these maxima should be visible both in
the QKR and the pendulum data. We decide to take the maximal
deviation within the first $10$ maxima. 


When requiring an accuracy of $10\%$ we can give as a boundary of validity
$\beta\gtrsim0.2$. In the case of the amplitude the criterion is not
as clear. Taking also the deviation in the amplitude into account we
concluded that $\beta\gtrsim0.3$ following this criterion. 

\subsection{Ensembles}
\label{ens_rotors}
In order to build ensembles we need to evaluate the integral in
\eref{eq:DefEnsemble}. This integral is approximated by a Riemann sum
with $N_\beta$ values of $\beta$ uniformly distributed in
$[\beta_1,\beta_1+\Delta\beta]$. 
It has been checked that $N_\beta$ should be of order of a few thousand
to get a reasonably good approximation for the integral in (\ref{eq:DefEnsemble}).


\begin{figure}
 \centering
 \includegraphics[scale=1]{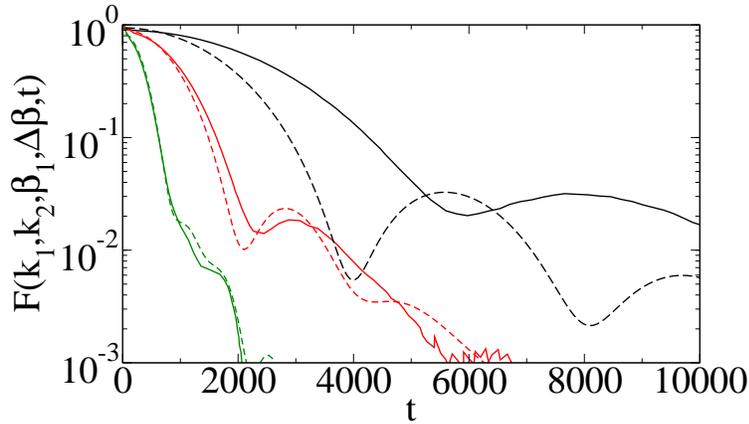}
 \caption{(Colour online) Comparison between the fidelity for the perturbative
   approach (dashed line) and the QKR (solid line), $\epsilon=0.05$,
   $k_1=0.6\pi$, $k_2=0.8\pi$, $\Delta\beta=0.06$, $\beta_1=0.06$
   (black, right), $\beta_1=0.14$ (red, middle), $\beta_1=0.22$ (green, left).} 
 \label{fig:ExplEnsTime}
\end{figure}

For the boundary of the phase space cell the same criterion as used
for single rotors is applied. In Fig.~\ref{fig:ExplEnsTime} a
few ensembles are shown. For the measure of correspondence we compare
the widths of the first few pseudo-oscillations and also demand a deviation
of less than $10\%$ here. The onset of the island behaviour near to a
resonance leads to small peaks in the fidelity as described in
\cite{Abb2009}. Therefore the occurrence of these peaks defines the
critical value near the resonance island. The intervals for several
$\epsilon$ are summarised in Table \ref{tab:range_of_validity}. The
upper bound is described quite well by the estimate due to the
pendulum approximation \eref{eq:Estimation_Island} and only fails for
the largest $\epsilon$ presented, i.e. far from the
semiclassical regime. 

\begin{table}
 \begin{tabular}{l|l|c}
  $\epsilon$&range for ensemble&theoretical upper bound\\ \hline
  $0.1$&$0.10-0.16<\beta_1,\beta_2<0.36-0.37$&$0.34$\\
  $0.05$&$0.12-0.16<\beta_1,\beta_2<0.39-0.41$&$0.39$\\
  $0.01$&$0.12-0.16<\beta_1,\beta_2<0.45-0.47$&$0.45$\\
 \end{tabular}  
\caption{Range of validity of the pendulum approximation. For
  simplicity we defined $\beta_2=\Delta\beta+\beta_1$.} 
  \label{tab:range_of_validity}
\end{table}

\section{Scaling of ensembles}

In this section we perform additional numerical analysis over a
larger range of parameters. This will confirm the range of validity of
our perturbative approach.
\begin{figure}
\centering
\includegraphics[clip]{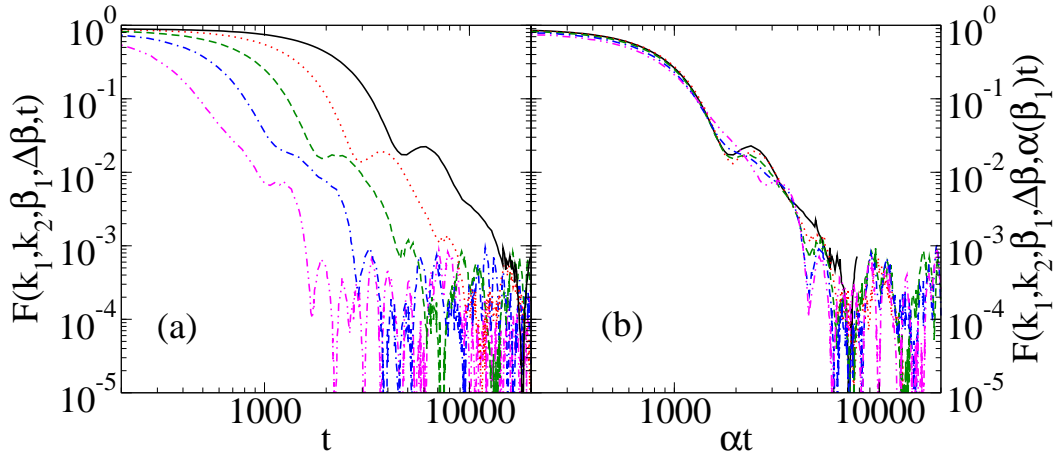}
 \caption{(Colour online) Fidelity of a few ensembles of $\beta-$rotors with
   $\epsilon=0.05$, $\Delta\beta=0.06$, $k_1=0.6\pi$, $k_2=0.8\pi$,
   $\beta_1=0.08$ (black \full), $\beta_1=0.12$ (red \dotted), $\beta_1=0.16$
   (green \dashed), $\beta_1=0.2$ (blue \chain) and $\beta_1=0.24$ (magenta \dashddot). In
   (a) the original data and in
   (b) the rescaled data are shown. The
   scaling factors are according to
   figure~\ref{fig:Scaling_factrors}.}
 \label{fig:ScalingExample}
\end{figure}

The ensembles in Fig.~\ref{fig:ScalingExample}a show
a similar fidelity as a function of time except for a shift along
the time axis. This suggests a 
rescaling of the time. First choose a reference value of $\beta$, say $\beta_\mathrm{ref}$. Then
we claim that we can map the fidelity for \emph{another}
$\beta_1$ on top of the reference value only by rescaling the
time. This can be more formally written in the following way
($\beta_\mathrm{ref}$ is chosen a priori):
\begin{equation}
  F(k_1,k_2,\beta_1,\Delta\beta, t)\approx
  F(k_1,k_2,\beta_\mathrm{ref},\Delta\beta,\alpha(\beta_1)^{-1}\ t)\ .
\label{scaling}
\end{equation}
The fidelity is scaled such that a box surrounding the mean decay can
be kept as narrow as possible. An example where the curves in
Fig.~\ref{fig:ScalingExample}a have been rescaled is
presented in Fig.~\ref{fig:ScalingExample}b. {One may notice that the formula (\ref{scaling}) comes from heuristic and
numerical observations. It is believed to work as long as the pendulum
approximation does for the kicked rotor. The latter
approximation cannot be controlled in a simple way \cite{Chirikov1979}. That is the reason
why we will only give here numerical bounds for the validity of
(\ref{scaling}).}

\begin{figure}
\begin{tabular}{cc}
 \subfigure[]{
 \includegraphics[clip=true]{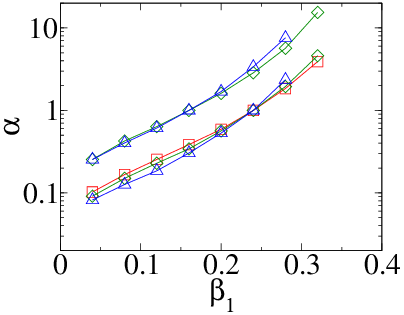}
  \label{subfig:Scaling_factor_dk02_eps_0005}
} &
 \subfigure[]{
 \includegraphics[clip=true]{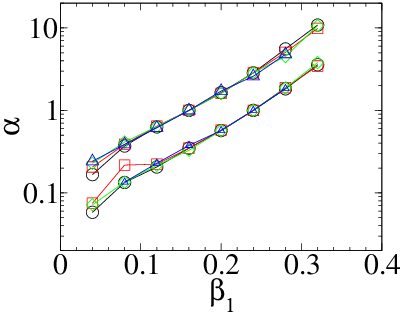}
  \label{subfig:Scaling_factor_dk01_eps_005}
} 
\end{tabular}
\caption{(Colour online) Scaling factors for several parameters. The scaling is done
  for two references $\beta_\mathrm{ref}=0.16$ (upper curves) and
  $\beta_\mathrm{ref}=0.24$ (lower curves). $k_2=0.8\pi$, $\Delta\beta=0.03$ (black
  circles), $\Delta\beta=0.06$ (red squares), $\Delta\beta=0.09$
  (green diamonds), $\Delta\beta=0.12$ (blue
  triangles). \subref{subfig:Scaling_factor_dk02_eps_0005} 
  $\epsilon=0.005$ $\Delta k=0.2\pi$,
  and
  \subref{subfig:Scaling_factor_dk01_eps_005} $\epsilon=0.05$ $\Delta
  k=0.1\pi$.}  
\label{fig:Scaling_factrors}
\end{figure}

This scaling procedure was done for several $\beta_1$, $\Delta\beta$, $\epsilon$ and
$k_1$, whereas $k_2$ was kept the same. The procedure was repeated for another reference
$\beta_\mathrm{ref}$. We expect the curves corresponding to different reference
values to be parallel, i.e. a different choice of $\beta_\mathrm{ref}$
just leads to a constant offset. The resulting plots are 
presented in Fig.~\ref{fig:Scaling_factrors}. We can see that the
scaling factor has no strong dependence on any of the parameters
shown. 

\begin{figure}
 \centering
  \includegraphics{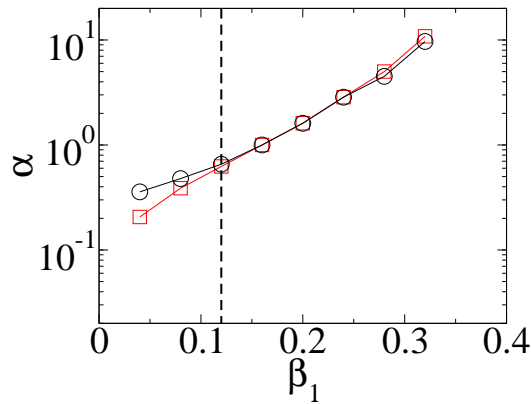}
  \caption{(Colour online) Scaling factors for the
    perturbative result (black circles)
    and QKR data (red squares), for $\epsilon=0.05$, $k_1=0.6\pi$,
    $k_2=0.8\pi$, $\beta_\mathrm{ref}=0.16$ and
    $\Delta\beta=0.06$. The dashed line is the border of
    correspondence between the QKR and the perturbative result given in
    Table~\ref{tab:range_of_validity}.}  
  \label{fig:Scaling_Comp_WKB_QKR}
\end{figure}

The same scaling procedure also has been performed for the
perturbative data computed from (\ref{Fidpend}). In Fig.~\ref{fig:Scaling_Comp_WKB_QKR} we show
the comparison between scaling factors of the perturbative result and
the QKR for one set of parameters. We observe a reasonable
agreement. Below $\beta_1\approx0.12$ there is a systematical
deviation between Eq.~\eref{Fidpend} and the QKR data. This
corresponds to the border of the phase space cell where we already
expect the pendulum approach to fail. 
One may conclude that the rescaling procedure is valid as long as
the pendulum approximation holds. More precisely in our case the range
of validity is given by the numerical bounds stated above in Sect.~4.2.

\section{Summary and outlook}

We have for the first time studied analytically the quantum fidelity
of initial conditions corresponding to rotational orbits in the
underlying pseudo-classical model. Using the pendulum approximation
for these orbits, we arrive at our main analytical result summarised
in Eq.~\eref{Fidpend}. Although we use a formally somewhat
inconsistent expansion in the perturbation parameter $\epsilon$, we
see that our approximation is rather good when including higher orders
in the dynamical phases, even if just the lowest order in the
amplitude of the wave functions is considered. We give clear ranges of
applicability of our approximation which were tested against numerical
simulations of the original quantum kicked rotor system. Within these
ranges a scaling hypothesis for the temporal decay of the fidelity is
found which is fulfilled by the original model as well as our
perturbative results. It would be interesting to set this scaling
hypothesis onto firm grounds by deriving it from first principles for
the here investigated rotational pseudo-classical orbits. This task is
left for future investigations. 

\ack
It is our great pleasure to thank Italo Guarneri for illuminating
discussions at the early stage of this work.
This work was supported by the DFG through FOR760, the Helmholtz
Alliance Program of the Helmholtz Association (contract HA-216
Extremes of Density and Temperature: Cosmic Matter in the
Laboratory), and within the framework of the Excellence Initiative
through the Heidelberg Graduate School of Fundamental Physics (grant
number GSC 129/1), the Frontier Innovation Fund and the Global
Networks Mobility Measures. 

\appendix
\section{Perturbative expansion of the pendulum energy levels}
\label{app_WKB}

We are interested in the quantum energy levels of the pendulum
Hamiltonian (\ref{Hpendulum}). It is worth reminding that
$\tilde{k}=k|\epsilon|$ where $|\epsilon|$ is our effective Planck
constant. We are interested in the regime of small $|\epsilon|$. 
Our procedure is the following: assume
first that $|\epsilon|$ is a fixed small quantity. Then write
perturbative expansions in $\tilde{k}$, which are 
valid whenever $\tilde{k}={\cal O}(|\epsilon|)$. At the very end we
will write more explicitly $\tilde{k}=k|\epsilon|$ to derive
$\epsilon-$semiclassical results. 

The classical action for the Hamiltonian (\ref{Hpendulum}) along a
trajectory from $\theta_i$ to $\theta$ at a fixed energy $E$ is:
\begin{equation}
  \label{Spend}
  S(\theta,\theta_i)=\int_{\theta_i}^\theta
  \sqrt{2(E-\tilde{k}\cos\varphi)}{\rm d}\varphi
  -\xi(\theta-\theta_i)\ ,
\end{equation}
where $\xi=\textrm{sgn}(\epsilon)(-\pi+\tau\beta)=\xi_0 + |\epsilon|\beta$.
The energy level can be well described using a WKB-like approach in the regime $|\epsilon|\to 0$ by:
\begin{equation}
  \label{quantpend}
  \int_{0}^{2\pi}
  \sqrt{2(E^k_m-\tilde{k}\cos\varphi)}{\rm d}\varphi -2\xi\pi=2\pi
  m|\epsilon|\ .
\end{equation}
Here the Maslov index is $0$ as the particle lives on a ring, hence
never meets any boundary. The quantization condition (\ref{quantpend}) with integers $m$
can be rewritten in a more efficient way as
\begin{equation}
  \label{quantpend2}
  4\sqrt{2(E_m^k-\tilde{k})}\
  \mathbb{E}\left(\rmi\sqrt{\frac{2\tilde{k}}{E_m^k-\tilde{k}}}\right)=
  2\pi\left[ (m+\beta)|\epsilon|+\xi_0\right]\ , 
\end{equation}
where $\mathbb{E}(\kappa)$ is the Legendre complete elliptic integral:
\begin{equation}
  \label{defE}
  \mathbb{E}(\kappa)=\int_0^{\pi/2} \sqrt{1-\kappa^2\sin(t)^2}{\rm
    d}t\ .
\end{equation}
Eq.~(\ref{quantpend2}) is the starting point of our perturbation
expansion. We assume from now that $\xi \ne 0$. For the left hand side
of (\ref{quantpend2}) the Taylor expansion of
$\mathbb{E}(\kappa)$ is used, see e.g. 8.114 (1) p.853 in \cite{Gradsteyn}:
\begin{equation}
  \label{TaylE}
  \mathbb{E}(\kappa)=\frac{\pi}{2}\left[ 1 -\frac{\kappa^2}{4}-
    \frac{3\kappa^4}{64}- \frac{5\kappa^6}{256} -
    \frac{175\kappa^8}{16384} + {\cal O}(\kappa^{10})\right]
\end{equation}
One gets: 
\begin{eqnarray}
   \mathbb{E}\left(\rmi\sqrt{\frac{2\tilde{k}}{E_m^k-\tilde{k}}}\right)=&
   \nonumber\\
  \frac{\pi}{2} \left[ 1 +
    \frac{\tilde{k}}{2E_m^k}  + \frac{5\tilde{k}^2}{16 E_m^k{}^2 } +
    \frac{9\tilde{k}^3}{32E_m^k {}^3} + \frac{241\tilde{k}^4}{1024E_m^k{}^4} 
+ {\cal O} \left(\frac{\tilde{k}^5}{E_m^k{}^5} \right) \right]&  \label{lhs}
\end{eqnarray}
The eigenenergy $E_m^k$ is assumed to have the following form:
\begin{equation}
  \label{Taylen}
  E_m^k=\frac{\xi_0^2}{2} +\xi_0(m+\beta) |\epsilon| +\alpha_2 \epsilon^2 +\alpha_3 |\epsilon|^3 +\alpha_4 \epsilon^4
\end{equation}
Identifying both parts of (\ref{quantpend2}) using (\ref{lhs}) leads
to the following results: 
\begin{eqnarray}
  \label{alpha2}
  \alpha_2&=&\frac{(m+\beta)^2}{2}+\frac{k^2}{4\xi_0^2}\\
  \label{alpha3}
  \alpha_3&=&-\frac{(m+\beta) k^2}{2\xi_0^3}\\
  \label{alpha4}
  \alpha_4&=&\frac{3}{4}\frac{(m+\beta)^2k^2}{\xi_0^4}+\frac{5}{64}\frac{k^4}{\xi_0^6}
\end{eqnarray}
Inserting (\ref{alpha2}), (\ref{alpha3}) and (\ref{alpha4}) into
(\ref{Taylen}) gives (\ref{E_pert}).

\end{document}